\documentclass[aps,prb,showpacs,twocolumn]{revtex4}
\usepackage{units}
\usepackage{amsfonts}
\usepackage{amssymb}
\usepackage{amsmath}
\usepackage{graphicx}
\usepackage{dcolumn}
\usepackage{bm}
\usepackage{color}

\begin{document}

\title{Nonlinear bias dependence of spin-transfer torque from atomic first
principles}
\author{Xingtao Jia$^1$, Ke Xia$^1$, Youqi Ke$^2$ and Hong Guo$^2$}
\affiliation{$^1$ Department of Physics, Beijing Normal University, Beijing 100875, China%
\\
$^2$ Centre for the Physics of Materials and Department of Physics, McGill
University, Montreal, PQ, H3A 2T8, Canada}
\date{\today }

\begin{abstract}
We report first-principles analysis on the bias dependence of spin-transfer
torque (STT) in Fe/MgO/Fe magnetic tunnel junctions. The in-plane STT
changes from linear to nonlinear dependence as the bias voltage is increased
from zero. The angle dependence of STT is symmetric at low bias but
asymmetric at high bias. The nonlinear behavior is marked by a threshold
point in the STT versus bias curve. The high-bias nonlinear STT is found to
be controlled by a resonant transmission channel in the anti-parallel
configuration of the magnetic moments. Disorder scattering due to oxygen
vacancies in MgO significantly changes the STT threshold bias.
\end{abstract}

\maketitle

\section{Introduction}

Magnetic tunnel junction (MTJ) has attracted great attention due to its
importance in magnetic random-access memory (MRAM) and read-sensor
technology. A basic MTJ is made of two ferromagnetic (FM) layers sandwiching
a thin insulating material. Digital information is coded by magnetic moments
of the FM layer being in parallel or anti-parallel configurations (PC or
APC). In commercial MRAMs, switching between PC and APC is achieved by
external magnetic fields. An emerging trend is to switch by spin-transfer
torque (STT) predicted theoretically\cite{Slonczewski1996} and demonstrated
experimentally\cite{Diao2005,Diao2007,Matsumoto2009} where a spin-polarized
current transfers angular momentum to a FM layer causing magnetic moment
reversal. STT holds great promise to simplify the MRAM structure, down-scale
device size, and reduce power consumption.

%[background on spin currents, non-equilibrium transport and SST]

One of the most important issues concerning STT is its dependence on
external bias voltage $V_b$ that drives the spin-polarized current in the
first place. Understanding this issue is rather difficult not only because
torque is a vector\cite{Sankey2008,Kubota2008,Deac2008,Oh2009,Wang2011} but
also because STT operates at non-equilibrium due to the flow of spin
currents. A physical picture concerning the bias dependence of STT, from the
linear to nonlinear bias regime, is yet to be established from atomic first
principles. Theoretically, $V_b$ dependence of STT was investigated by model
analysis such as the tight-binding model\cite{Theodonis2006,Kalitsov2009}
and free-electron model.\cite{Xiao2008,Wilczynski2008,Manchon2008}
%,Wilcinsky2011}.
More recently, Heiliger and Stiles\cite{Heiliger2008} analyzed STT for an
MgO based MTJ based on equilibrium electronic structure obtained from
density-functional theory (DFT). The in-plane STT was calculated\cite%
{Heiliger2008} up to $V_b=\pm 0.5$ V and showed essentially a linear
dependence on $V_b$, and the out-of-plane STT was found to be quadratic in $%
V_b$, which is consistent with tight-binding\cite{Theodonis2006} and
experimental results.\cite{Sankey2008,Wang2009}

Existing calculations provided valuable understanding of STT at the small $%
V_{b}$ regime. To establish a complete picture, it is important to
investigate STT at higher bias from atomic first principles. Indeed, recent
experimental data\cite{Wang2011} already showed a strong nonlinear $V_{b}$
dependence of in-plane STT, and observable non-quadratic out-of-plane STT
beyond $V_{b}\approx 0.2$V. It is the purpose of this work to report a
first-principles analysis of STT in the most technologically important
Fe/MgO/Fe MTJ from low- to high-bias regimes, focusing on its nonlinear $%
V_{b}$ dependence. We found that the in-plane STT changes from linear to
nonlinear dependence as $V_{b}$ is increased from zero. The angle dependence
of STT is symmetric at low bias but asymmetric at high bias. The nonlinear
behavior is marked by a threshold point in the STT versus bias curve which
is controlled by a resonant transmission channel in the APC current. We also
found that the nonlinear STT is affected by atomic defects$-$oxygen
vacancies in MgO$-$hence, in principle, can be tuned by interface
engineering. Results are compared to the corresponding experimental data.

This paper is organized as follows. In Sec. II, we give the details of our
calculation based on Keldysh nonequilibrium Green's function (NEGF) and
wave-function-matching function method. In Sec. III, we present our results
on Fe/MgO/Fe(001) MTJs with clean and disordered interfaces. Section IV is
our summary.

\section{Electronic structure and transport calculation}

\begin{figure}[tbp]
\includegraphics[width=8.5cm]{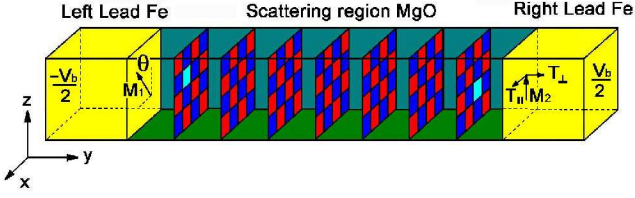}
\caption{(Color online) Sketch of a Fe/MgO/Fe(001) MTJ having seven MgO
layers. The magnetization of the left Fe lead ($\mathbf{M}_1$) is fixed and
that of the right Fe lead ($\mathbf{M}_2$) is free to rotate. $\mathbf{M}%
_{2} $ is pointing along the $\mathit{z}$ axis; $\mathbf{M}_{1}$ lies in the
\textit{x-z} plane, the angle $\protect\theta $ is the angle between $%
\mathbf{M}_1$ and $\mathbf{M}_2$. The red and blue grids in the scattering
region (MgO) denote the O and Mg atoms, respectively. There may be some
interfacial disorder (oxygen vacancy, cyan grid) at the Fe/MgO interfaces.
The applied bias eV$_{b}$=$\protect\mu_{R}-\protect\mu _{L}$, where $\protect%
\mu_{R}$ and $\protect\mu_{L}$ are chemical potentials of the right and left
leads.}
\label{fig1}
\end{figure}

The MTJ we consider consists of an MgO barrier sandwiched by two
semi-infinite Fe leads shown in Fig.\ref{fig1}. The system is periodic in
the $x$-$z$ plane and current flows along $y$ corresponding to the (001)
material-growth direction. A very small lattice mismatch between Fe and MgO
is neglected by fixing the interfacial atoms at their bulk bcc positions.
Our calculation is from first principles where DFT is carried out within the
Keldysh nonequilibrium Green's function (NEGF) method.\cite{jeremy,ke08}
Here, DFT determines the Hamiltonian of the open device structure, and NEGF
determines the nonequilibrium statistics of the device operation and the
nonequilibrium density matrix. The NEGF-DFT is solved self-consistently
under finite bias during current flow. For devices having oxygen vacancies
(OV, the most energetic favorable defects), we further apply the
nonequilibrium-vertex-correction (NVC) theory\cite{ke08} for disorder
averaging at both the nonequilibrium-density-matrix level and
transport-calculation level. The NEGF-DFT-NVC formalism allows one to
self-consistently calculate quantum transport properties from atomic first
principles without phenomenological parameters. We refer interested readers
to the original work.\cite{ke08}

Moreover, The self-consistent nonequilibrium potentials were used as input
to a TB-MTO wave-function-matching calculation. The scattering wave
functions of the whole system were obtained explicitly. Because we cannot
embody the nonequilibrium NVC into the out-of-plane STT calculations at
present, a supercell is used to assess out-of-plane STT in the presence of
interfacial disorder. In the calculations, an $800\times 800$ \textit{k}%
-mesh is used to sample the two-dimensional Brillouin zone (BZ) in the $x$-$%
z $ plane (see Fig. \ref{fig1}) to ensure accurate convergence. Other
numerical details are similar to those of Ref. \onlinecite{Wang2008} and %
\onlinecite{ke10}.

To check our electronic structure of Fe/MgO/Fe, we calculate the zero-bias
TMR for 7 monolayers (ML) MgO barrier, which is 5100\% with the perfect
Fe/MgO interface while decreases dramatically to 350\% (78\%) with 3\% (7\%)
OV at both interfaces. The TMR of a clean junction is consistent with the
published first-principles calculation,\cite{Butler2001} while the TMR of
junctions with vacancies are in the range of experimental measurement.\cite%
{YUASA2004,Faure-Vincent2003,Parkin2004} The results indicate that we have
got the right electronic structure.

\section{Bias dependence of STT}

\begin{figure}[tbp]
\includegraphics[width=8.5cm]{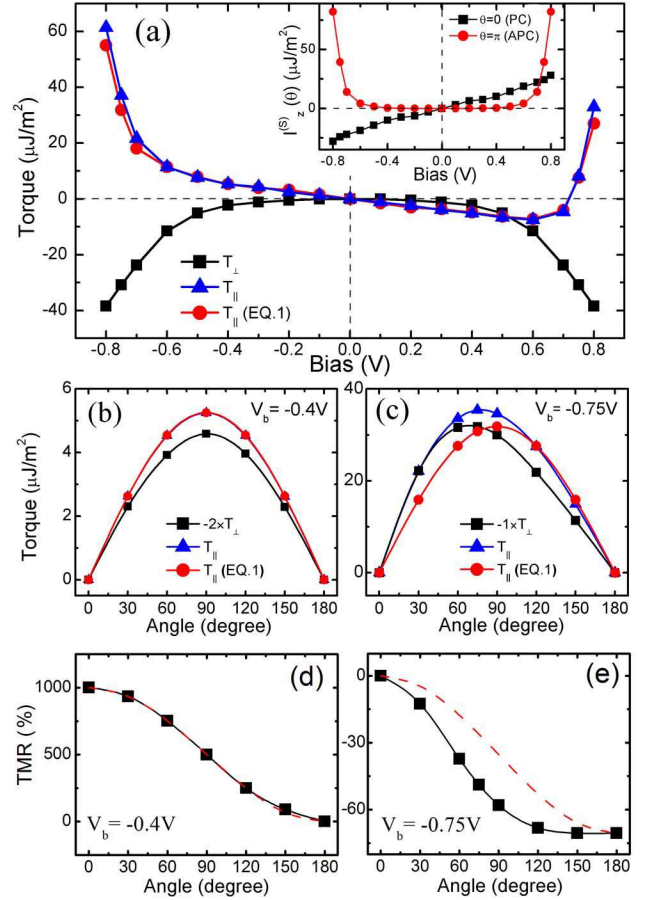}
\caption{(Color online) (a) STT of ideal Fe/MgO(7L)/Fe with angle $\protect%
\theta =90^{o}$ between \textbf{M}$_{1}$ and \textbf{M}$_{2}$. Black squares
and red circles are out-of-plane and in-plane STT calculated by the
wave-function-matching method, respectively, the blue up-triangle is the
in-plane STT calculated by Eq. (1). Inset: spin currents in PC and APC. (b)
and (c): angular dependence of STT at $V_{b}=-0.4$, and $-0.75$ V,
respectively. (d) and (e): angular dependence of TMR at $V_{b}=-0.4$, and$%
-0.75$ V, respectively; the red dash lines are standard cosine-dependence
TMR according to Ref. \onlinecite{Slonczewski1989}. }
\label{fig2}
\end{figure}

Before presenting results, let's consider some general trends. Neglecting
any spin dependent scattering in the tunnel barrier, the in-plane STT can be
obtained from conservation of spin current,\cite{Theodonis2006}
\begin{equation}
\mathbf{T}_{||}(\theta )=\frac{I_{z}^{(s)}(\pi )-I_{z}^{(s)}(0)}{2}\mathbf{M}%
_{\mathbf{2}}\times (\mathbf{M}_{\mathbf{1}}\times \mathbf{M}_{\mathbf{2}})
\label{Torque_theory}
\end{equation}%
where $\theta $ is the relative angle between magnetization $\mathbf{M}_{1}$
and $\mathbf{M}_{2}$ (see Fig. \ref{fig1}). $I_{z}^{(s)}(0)\equiv \frac{%
\hslash }{2e}[I^{\uparrow }(0)-I^{\downarrow }(0)]$ is the spin-polarized
current density in PC and $I_{z}^{(s)}(\pi )\equiv \frac{\hslash }{2e}%
[I^{\uparrow }(\pi )-I^{\downarrow }(\pi )]$ is that of APC. By structural
symmetry, $I_{z}^{(s)}(0)$ should be an odd function and $I_{z}^{(s)}(\pi )$
an even function of $V_{b}$. Hence, at not too large $V_{b}$, we can write $%
I_{z}^{(s)}(0)\approx \gamma _{1}V_{b}$ and $I_{z}^{(s)}(\pi )\approx \gamma
_{2}V_{b}^{2}$, where $\gamma _{1}$ and $\gamma _{2}$ are constants. From
these considerations, Eq. (\ref{Torque_theory}) suggests $I_{z}^{(s)}(0)$ to
dominate $T_{||}$ at small $V_{b}$ (positive $V_{b}$) while $I_{z}^{(s)}(\pi
)$ to dominate at higher $V_{b}$. Hence, $T_{||}$ should start from a linear
dependence and turn into a nonlinear dependence as $V_{b}$ is increased; the
turning point defines a \textquotedblleft threshold" bias whose value is
determined roughly by the material-specific constants $\gamma _{1,2}$. At
low bias, the out-of-plane STT demonstrates a simple quadratic relation\cite%
{Theodonis2006} that we can write as $T_{\perp }=\gamma _{\perp }V_{b}^{2}$.

After the NEGF-DFT self-consistent calculation is converged,\cite{ke08} spin
currents $I^{\uparrow }$ and $I^{\downarrow }$ can be calculated by NEGF\cite%
{ke08} and the in-plane STT $T_{||}$ obtained from Eq. (\ref{Torque_theory}%
). The out-of-plane STT $T_{\perp }$ (also $T_{||}$) can be obtained from a
scattering-wave-function approach following Ref. \onlinecite{Wang2008}.
Figure \ref{fig2} shows plots of the calculated $T_{||}$ and $T_{\perp }$
versus $V_{b}$ for a perfect Fe/MgO/Fe MTJ (no disorder) having seven layers
of MgO, where magnetization $\mathbf{M}_{1}\perp \mathbf{M}_{2}$, \emph{i.e.,%
} $\theta =90^{o}$ (see Fig. \ref{fig1}). Seven-layer MgO is used because it
is the thickness of experimental device\cite{Wang2009} with which we shall
compare results (see below). The blue (up-triangle, $T_{||}$) and black
(solid square, $T_{\perp }$) curves were calculated from the scattering wave
functions.\cite{Wang2008} The red curve (solid circle, $T_{||}$) was
calculated from Eq. (\ref{Torque_theory}), which agrees with the blue curve
very well at least for lower $V_{b}$.\cite{foot1} At lower bias up to $\pm
0.5$V, $T_{||}$ is linear in $V_{b}$ and it is in agreement with Ref.%
\onlinecite{Heiliger2008}. As $V_{b}$ increases, $T_{||}$ becomes gradually
nonlinear and a clear threshold at $V_{b}\sim 0.65$V (e.g., turning point of
the curves) is seen, as expected from the general considerations discussed
above.

The inset of Fig. \ref{fig2}(a) plots the spin-current density versus $V_{b}$%
. Spin-polarized current density $I_{z}^{(s)}(0)$ is indeed linear in $V_{b}$
whose slope is $\gamma _{1}\approx 35\mu $JV$^{-1}$m$^{-2}$. Spin-polarized
current density $I_{z}^{(s)}(\pi )$ is, as expected, an even function of $%
V_{b}$ and, by fitting to a quadratic form $\gamma _{2}V_{b}^{2}$, we found
that $\gamma _{2}\approx 5\mu $JV$^{-2}$m$^{-2}$ for $V_{b}\leq 0.5$ V. The
small value of $\gamma _{2}$ makes $I_{z}^{(s)}(\pi )$ to have a very weak $%
V_{b}$ dependence, until its sudden increase at $V_{b}\sim 0.6$ V, resulting
in the threshold of $T_{||}$ at $\sim $0.65 V.

The out-of-plane STT shows quadratic behavior at the lower bias with $\gamma
_{\perp }\approx -13\mu $JV$^{-2}$m$^{-2}$, and it is very close to the
published calculated value of $-14\mu $JV$^{-2}$m$^{-2}$ with a six-ML MgO
barrier.\cite{Heiliger2008}
%, and experimental $%-11$ and $-5\mu $JV$^{-2}$m$^{-2}$ with a 1.0\cite{Kubota2008} and 1.25\cite%
%{Wang2009} nm MgO barrier, respectively), which departs from this
%behavior at higher bias, with threshold V$_{b}\sim 0.5$V, as shown
%by the solid squares in Fig.\ref{fig2}(a).}
Quantitatively, $T_{\perp }<T_{||}$ for $V_{b}<0.5$ V but reaches a similar
scale when $V_{b}>0.5$ V.

Not only the bias dependence of STT [see Fig. \ref{fig2}(a)] is changed by
the sudden increase of APC current $I_{z}^{(s)}(\pi )$, the angular
dependence of STT is also modified. As shown in Fig. \ref{fig2}(b), the
angular dependence behaves as sin$\theta $ at a small bias $V_{b}=-0.4$ V.
It however deviates from the sine dependence at a higher bias $V_{b}=-0.75$
V [see Fig. \ref{fig2}(c)]. Experimental evidence of asymmetric
angular-dependent STT at higher bias had been already observed before.\cite%
{Wang2009}

Can the angular dependence of in-plane STT be understood by the angular
dependence of the charge current or TMR? We give the angular dependence of
TMR at a small bias $V_{b}=-0.4$ V [see Fig. \ref{fig2}(d)], and a large
bias $V_{b}=-0.75$ V [see Fig.\ref{fig2}(e)]. The angular dependence of TMR
at low bias follows the cosine function\cite{Slonczewski1989} and deviates
from the cosine function at high bias. At high bias, the in-plane STT shows
an asymmetry and the peak is close to the parallel side, while the
angular-resolved TMR decreases always faster than cosine function as the
magnetization goes from parallel to antiparallel. Notice here, we have
negative TMR at high bias. So, the angular dependence of TMR cannot explain
our calculated angular dependence of in-plane STT.

\begin{figure}[tbp]
\includegraphics[width=8.5cm]{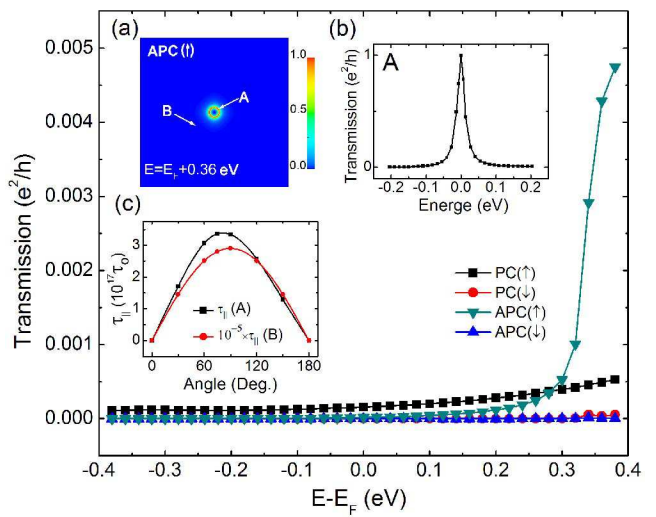}
\caption{(Color online) Energy dependent spin polarized transmission
coefficient of ideal Fe/MgO(7L)/Fe MTJ at $V_{b}=\pm $0.75V. Insert (a): the
$\mathbf{k}_{||}$ resolved transmission hot spot of the $\uparrow $ -channel
in APC at energy E=E$_{F}$+0.36eV. Insert (b): the energy dependent
transmission of a bright point A on the ring of the insert (a) at k$_{x}$=k$%
_{y}$=0.278\AA $^{-1}$. Insert (c): angular dependent in-plane STT at
resonant point A k$_{x}$=k$_{y}$= 0.278\AA $^{-1}$, and normal point B k$%
_{x} $=k$_{y}$= -1.624\AA $^{-1}$ of the insert (a); therein, $\protect\tau%
_{o}\equiv \frac{\hslash }{2e}k\Omega ^{-1}m^{-2}$ sets the unit\protect\cite%
{foot2}}
\label{fig3}
\end{figure}

It is the sudden increase of APC current $I_{z}^{(s)}(\pi )$ at a higher
bias $V_{b}=0.6$ V [inset of Fig. \ref{fig2}(a)] that led to the nonlinear
bias dependence of STT. To understand the microscopic origin of this sudden
increase, we plot spin-resolved transmission coefficient versus electron
energy $E$ (relative to the Fermi energy $E_{F}$) in Fig. \ref{fig3},
calculated at $V_{b}=0.75$V. The $\uparrow $ or $\downarrow $ denotes the
majority or minority spin channel in the left Fe lead. For APC, the $%
\uparrow $ channel of the left lead transmits to the $\downarrow $ channel
of the right lead through the MgO. Figure .\ref{fig3} shows that there is an
abrupt increase in $\uparrow $ channel transmission at about $E=0.3$ eV. For
a symmetric MTJ, roughly half $V_{b}$ is dropped at each Fe/MgO interface.
Hence the abrupt increase of transmission at $E\approx 0.3$ eV should
contribute to APC current when the bias $V_{b}=0.6$ V. This is indeed what
is found [inset of Fig .\ref{fig2}(a)]. We conclude that the sudden increase
of $I_{z}^{(s)}(\pi )$ is due to the abrupt increase of the $\uparrow $
channel transmission, which, importantly, is also due to bias-dependent
transport features as we explain now.

The inset (a) of Fig.\ref{fig3} plots the transmission hot spot of APC($%
\uparrow $) in the two-dimentional (2D) BZ, which is the transverse momentum
($\mathbf{k}_{||}$) resolved transmission coefficient. There is a clear
bright ring surrounding the $\Gamma $ point. The abrupt increase of the APC $%
\uparrow $ channel transmission is due to this bright ring. Focusing on one
particular $k$ point on this ring and scanning the electron energy, a single
resonant peak is discovered in the transmission as shown in the inset (b) of
Fig. \ref{fig3}.

Importantly, this sharp resonance is different from the well-known resonance
states on Fe surfaces.\cite{Xu2006} In particular, we found that this sharp
APC resonance is only established at high bias and is located between the $%
\Delta _{1}$ band in the left Fe and the $\Delta _{5},\Delta _{2\prime }$
bands in the right Fe. The contribution of $\Delta _{5}$ and $\Delta
_{2\prime }$ bands to the total transmission is k$_{||}$ dependent, giving
rise to the ring in the hot spot. A higher bias significantly changes the
shape of the tunnel barrier: we checked that by artificially flattening the
barrier potential from the biased slop shape, this sharp APC resonant
channel disappears.

Actually, the existence of resonance is not only responsible for the sudden
increase of STT but also responsible for the deviation of angular-dependent
STT as shown in the inset (c) of Fig. \ref{fig3}. When at a resonant ring
such as A point ($k_{x}=k_{y}=0.278$ \r{A}$^{-1}$), the angular-dependent
in-plane STT deviates from a sine function and follows similar behavior as
that in Fig. \ref{fig2}(c). When far from the resonant ring such as B point (%
$k_{x}=k_{y}=-1.624$ \r{A}$^{-1}$), the angular dependent in-plane STT
follows a sine function as well. Under most situations, at high bias, the
resonance dominates the in-plane STT. Without resonant states, the Bloch
wave should decay very fast in the tunnel barrier, the multiple scattering
should be negligible, and Eq. \ref{Torque_theory} should hold in this case.
However, in the presence of resonant states, the wave function of these
states will decay little in the barrier and Eq. \ref{Torque_theory} is not
valid anymore.

\begin{figure}[tbp]
\includegraphics[width=8.5cm]{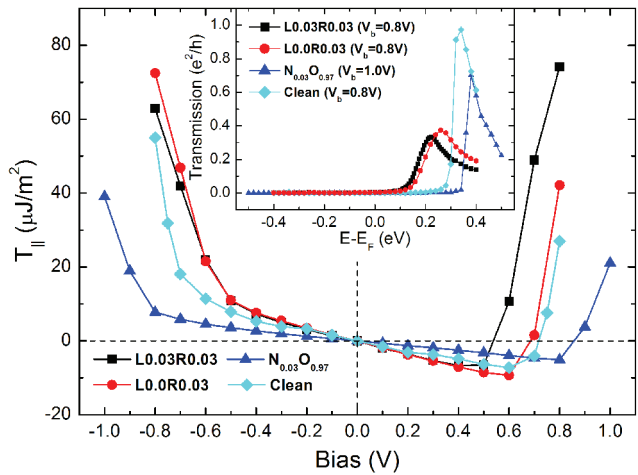}
\caption{(Color online) STT of Fe/MgO(7L)/Fe with oxygen vacancy
disorder versus bias. LxRy denotes that there are $x$ and $y$ OVs at
the left and right interface, respectively. N$_{0.03}$O$_{0.97}$
denotes that $3\%$ oxygen atoms at both interfaces are replaced by
nitrogen atoms.} \label{fig4}
\end{figure}

Having understood the microscopic physics behind the nonlinear bias
dependence of STT, we note that the obtained threshold bias at $V_{b}\sim 0.6
$ V for the \emph{ideal} junction is higher than that reported
experimentally,\cite{Kubota2008} which was about $0.3$ V. Since experimental
devices are never ideal, an investigation of defect scattering and its
effect on STT is warranted. There are several possible defects such as
oxygen vacancy in MgO, interfacial roughness,\cite{Heiliger2008} material
imperfections, etc. It was reported experimentally that due to compressive
strain during crystal growth, oxygen vacancies (OV) are accommodated in MgO.%
\cite{Mather2006} Both theory\cite{ke10} and experiment\cite{Miao2008} have
shown that existence of OV can drastically influence spin-polarized
transport and its effect on STT is therefore expected.

We calculated the in-plane STT from Eq. (\ref{Torque_theory}) by putting OVs
in the interfacial MgO layer immediately adjacent to Fe. An alloy model O$%
_{1-x}$Va$_{x}$ is used, where Va stands for vacancy and $x$ is the OV
concentration. The disorder average is carried out by the NVC theory\cite%
{ke08} mentioned above. Results are plotted in Fig. \ref{fig4} versus $V_{b}$%
, and several observations are in order.

First, at low $V_{b}$, $T_{||}$ is linear for both disordered ($x\neq 0$)
and clean samples ($x=0$)---hence, the linear regime is not qualitatively
altered by OVs. Second, both symmetric junctions where the same OV
concentration is on both Fe/MgO interfaces (black curve with solid squares),
and asymmetric junctions where OVs only exist at one Fe/MgO interface (red
curve with solid circles), yield almost the same $T_{||}$. In particular,
its slope $\tau _{||}\equiv dT_{||}/dV\approx -5.5\times 10^{13}\tau _{o}$.
%where $\tau _{o}\equiv \frac{\hslash }{2e}k\Omega
%^{-1}m^{-2}$ sets the unit\cite{foot2}.
This value is only slightly larger than that of the ideal junction, which
has $\tau _{||}=-4.7\times 10^{13}\tau _{o}$. Hence, a small amount of OV
(e.g. 3\%) does not alter $T_{||}$ significantly. These results should be
compared to the experimental value of the slope $\tau _{||}=-3.2\times
10^{13}\tau _{o}$ measured at $V_{b}=\pm 0.3$ V.\cite{Wang2009} Giving the
uncertainties in comparing our atomic structure to the experimental one, the
quantitative consistency of our results is quite satisfactory. Third, in
Ref. \onlinecite{ke10}, it was shown that filling the OVs with nitrogen
atoms can significantly reduce disorder-induced diffusive scattering and
drastically increase the TMR ratio. Here, filling the OVs with N using the
alloy model O$_{1-x}$N$_{x}$, we found that the $T_{||}$ is somewhat
decreased, namely, its slope $\tau _{||}=-1.9\times 10^{13}\tau _{o}$ (blue
curve with up-triangles).

Similar to the ideal junction, systems with OV also show abrupt increase of $%
T_{||}$ at higher bias, which, we found, is also due to the quantum
resonance in the 2D BZ discussed above. Different from the ideal case,
existence of OVs shifts the threshold bias of $T_{||}$ to lower values thus
bringing it closer to the experimental result.\cite{Kubota2008} As shown in
the inset of Fig. \ref{fig4}, the effect of OV on $T_{||}$ can also be
understood by investigating transmission of the $\uparrow $ channel in APC.

The OVs induce two new effects. First, interfacial OVs broaden the area
having high transmission values in the hot spots and also suppress the
transmission peak from 1.0 $e^{2}/h$ to about 0.35 $e^{2}/h$ (at 3\% OV on
both interfaces). The overall effect is to enhance $T_{||}$ slightly.
Second, interfacial OVs shift the resonant peak in the $\uparrow $ channel
of APC to lower energy and, as a result, the threshold bias is now shifted
to lower values.

Remarkably, from Fig. \ref{fig4} one can determine which Fe/MgO interface
dominates STT. When OVs exist at both interfaces (black curve with solid
squares), the threshold bias shifts to lower values from those of the clean
sample for both $V_{b}$ polarity. When OVs exist only at the right Fe/MgO
interface (red curve with solid circles), the threshold bias shifts to lower
value only in the negative $V_{b}$. When OVs are filled by N atoms (blue
curve with up-triangles), the threshold bias shifts to higher $V_{b}$ for
both polarities.

Quantitatively, the threshold bias is $\sim $0.45V in the presence of 3\% OV
at interface. It is consistent with the experimental value $\sim $0.4 V (see
Ref. \cite{Wang2009}) with a 1.25-nm (6-ML) MgO barrier. A more recent
experiment measured a threshold bias $\sim $0.2 V (see Ref. \cite{Wang2011})
with a 1.0-nm (5-ML) MgO barrier. The experimentally measured resistance is
1.5 $\Omega \mu $m$^{2}$ also close to our calculated value for a junction
with a 5-ML MgO barrier.

The effects of OVs on the out-of-plane STT are also very important. We have
examined a junction using the wave-function-matching method. The calculation
was done in a $7\times 7$ lateral supercell containing five randomly
distributed OVs at each Fe-MgO interface (thus $x\sim 10\%$). For $V_{b}=0.5$
V and $E=E_{F}+0.25$ eV, the calculated in-plane and out-of-plane STT are $%
20.30\times 10^{14}\tau _{o}$ and $7.32\times 10^{14}\tau _{o}$,
respectively. These values are larger than that of the ideal junction which
are $1.06\times 10^{14}\tau _{o}$ and $0.94\times 10^{14}\tau _{o}$ for
in-plane and out-of-plane STT, respectively. The larger in-plane and
out-of-plane STT in the 10\% OV sample is attributed to the resonant
tunneling discussed above. These results indicate that the interfacial OV
disorder is not detrimental to the value of the out-of-plane STT. In
comparison, interface roughness disorder quenches the out-of-plane STT as
reported in Ref. \onlinecite{Heiliger2008}.

\section{Summary}

In summary, the bias dependence of STT was investigated from atomic first
principles, covering both low- and high-bias regimes. STT shows weak and
linear bias dependence at small $V_{b}$ but strong and nonlinear dependence
at higher $V_{b}$. The angle dependence of STT is symmetric at low bias but
asymmetric at high bias. The nonlinear behavior is marked by a threshold
bias in the STT versus bias curve and is controlled by a resonant
transmission channel in APC. Very importantly, disorder scattering by small
amount of oxygen vacancies in MgO lowers the STT threshold bias but not the
STT value, suggesting that the nonlinear STT can be tuned by clever
interfacial engineering as also reported in Ref. \onlinecite{Miao2008} where
crystal-growth temperature changes the OV contents.

\section{Acknowledgements}

K.X. thanks financial support by the National Basic Research Program of
China (973 Program) under grant No. 2011CB921803 and NSF-China grant No.
10634070; H.G. thanks NSERC of Canada, FQRNT of Quebec, and CIFAR.


\begin{thebibliography}{99}
\bibitem{foot1} There is about 20\% difference in the in-plane STT between
the two methods at higher bias 0.8V.
\end{thebibliography}
\end{document}